\documentclass[aps,prb,twocolumn,superscriptaddress]{revtex4}
\usepackage{epsfig}
\usepackage{graphicx}
\usepackage{amsmath}
\usepackage{amsfonts}
\usepackage{amssymb}
\usepackage{bm}
\usepackage{bbold}
\usepackage{epsfig}
\usepackage{graphicx}
\usepackage{color}
\usepackage{pictex}
\usepackage[english]{babel}
\usepackage{mathtools}
\begin{document}
\title{String and conventional order parameters in the solvable modulated quantum chain}
\author{Gennady Y. Chitov}
\affiliation{Department of Physics, Laurentian University, Sudbury, Ontario,
P3E 2C6 Canada}
\author{Toplal Pandey}
\affiliation{Department of Physics, Laurentian University, Sudbury, Ontario,
P3E 2C6 Canada}
\author{P. N. Timonin}
\affiliation{Physics Research Institute,
Southern Federal University, 194 Stachki ave.,  Rostov-on-Don, 344090 Russia}
\date{\today}

%
%xxxxxxxxxxxxxxxxxxxxxxxxxxxxxxxxxxxxxxxxxxxxxxxxxxxxxxxxxxxxxxxxxxxxxxxxxxxxxx
%
\begin{abstract}
The phase diagram and the order parameters of the exactly solvable quantum $1D$ model are analysed.
The model in its spin representation is the dimerized $XY$ spin chain in the presence of uniform
and staggered transverse fields. In the fermionic representation this model is the dimerized noninteracting 
Kitaev chain with a modulated chemical potential.  The model has a rich phase diagram which contains
phases with local and nonlocal (string) orders. We have calculated within the same systematic framework
the local order parameters (spontaneous magnetization) and the nonlocal string order parameters, along with
the topological winding numbers for all domains of the phase diagram. The topologically nontrivial phase is
shown to have a peculiar oscillating string order with the wavenumber $q=\pi/2$, awaiting for its experimental confirmation.
\end{abstract}
\maketitle

%
%
%%%%%%%%%%%%%%%%%%%%%%%%%%%%%%%%%%%%%%%%%%%%%%%%%%%%%%%%%%%%%%%%%%%%%%%%%%%%%%
%
\section{Introduction}\label{Intro}
%
%%%%%%%%%%%%%%%%%%%%%%%%%%%%%%%%%%%%%%%%%%%%%%%%%%%%%%%%%%%%%%%%%%%%%%%%%%%%%%
%
%
%
According to the Landau theory, phases are distinguished by different types of long-ranged order, or its absence. The order is described
by an appropriately chosen order parameter, understood implicitly as a local quantity. \cite{Landau5}
There is a quite large number of examples of low-dimensional fermionic or spin systems as chains, ladders,
frustrated magnets, topological and Mott insulators, etc, \cite{FradkinBook13,TI,Ryu10,Montorsi12,HiddenSSB,SOPladders,Kitaev06,Delgado,Kim,UsLadd,KitHeis2019} (and more references in there) which clearly manifest distinct phases, criticality, but lack conventional local order even at zero temperature.

In the related recent work we have systematically demonstrated how the Landau formalism can be extended to deal with nonconventional quantum orders. \cite{GT2017,GYC2018} The key point is to incorporate nonlocal string operators, \cite{denNijs89}
string correlation functions, and string order parameters (SOPs). The appearance of nonlocal SOP is accompanied
by a hidden symmetry breaking. \cite{HiddenSSB} The local and nonlocal order parameters are related by duality, \cite{GT2017,GYC2018,Kogut79,ChenHu07,Xiang07,Nussinov} and it is eventually a matter of choice of variables of the Hamiltonian.

There are some additional aspects of quantum order quantified by, e.g., winding or Chern numbers, Berry phases, concurrence,
entanglement, \cite{FradkinBook13} which are not reducible to the parameters of the conventional Landau theory. These quantities provide
rather complementary description and do not seem to be indispensable. \cite{GT2017,GYC2018}

Nonlocal order parameters are known to be instrumental to probe hidden orders in various low-dimensional systems
\cite{SOPladders,Delgado,Kim,KitHeis2019,GT2017,GYC2018,ChenHu07,Xiang07,Nussinov,Berg08,Rath13}.
A SOP naturally becomes a part of the Landau paradigm, since its critical index $\beta$ satisfies the standard scaling relations
known for conventional order parameters \cite{GT2017,GYC2018}, and thus can be used to determine the universality class of a given
transition. The challenges in dealing with SOP are two-fold: from the theoretical side, in most of cases, this parameter quantitatively
can be obtained via some arduous simulations. Experimentally, the string correlation functions are notoriously hard to measure. However,
in the light of recent reports on experimental observation of bosonic string order, \cite{Enders11} one can expect more progress in
observation of SOPs in the near future.

In the context of said hurdles, it is really important to gain more insight on SOPs by dealing with rather simple but nontrivial models.
The model we study, in the guise of the dimerized $XY$ chain with homogeneous and alternating transverse fields has been known for several decades. \cite{Perk75}  It is exactly solvable, and its spectrum and phase diagram are well known. \cite{Perk75,TIMbook} However, the explicit calculations for the spontaneous magnetization seem to be missing in the literature. More importantly, the nature of the order in the topological phase was not clarified, and here we report our finding of its nonlocal string order, modulated with the wavevector $q=\pi/2$. The model in its fermionic representation is the Kitaev chain \cite{Kitaev2001} of noninteracting fermions with dimerized hopping and modulated (chemical) potential. The solvable Kitaev models with dimerizations and spatial modulations of potential were studied very actively in recent years with the focus on their topological phases with hidden orders and Majorana edge states
\cite{DeGottardi11,Lang12,Cai,EzawaNagaosa14,Zeng16,Miao17a,Ezawa17,Miao17b,Ohta16,Ghadimi17,Katsura,Monthus18,Wang18,GYC2018} In the context of current research interest, the present model provides a nice exactly solvable example with rich critical properties, when the phase diagram contains both conventional local and quite peculiar nonlocal orders.

The rest of the paper is organized as follows: In Sec.~\ref{Model} we introduce the spin and fermionic
representations of the model. We also discuss its spectrum, phase diagram, and the field-induced magnetization.
Sec.~\ref{OPs} contains the results. We present the formalism, the calculation of the spontaneous magnetization
(local order parameter) for the magnetic phase, the string order parameter for the topological phase, and  the winding numbers.
The Appendices contain details on the Majorana representations of the Hamiltonian and additional technical information on the
string operators and string order parameters. The results are summarized and discussed in the
concluding Sec.~\ref{Concl}.
%
%
%
%xxxxxxxxxxxxxxxxxxxxxxxxxxxxxxxxxxxxxxxxxxxxxxxxxxxxxxxxxxxxxxxxxxxxxxxxxxxxxx
%
\section{Model}\label{Model}
%
%xxxxxxxxxxxxxxxxxxxxxxxxxxxxxxxxxxxxxxxxxxxxxxxxxxxxxxxxxxxxxxxxxxxxxxxxxxxxxx
%
%
%
%
%xxxxxxxxxxxxxxxxxxxxxxxxxxxxxxxxxxxxxxxxxxxxxxxxxxxxxxxxxxxxxxxxxxxxxxxxxxxxxx
%
\subsection{Spin and fermionic representations of the model}
%
%xxxxxxxxxxxxxxxxxxxxxxxxxxxxxxxxxxxxxxxxxxxxxxxxxxxxxxxxxxxxxxxxxxxxxxxxxxxxxx
%
In this subsection we define the model and recapitulate its main results known explicitly or implicitly from earlier work.
The spin Hamiltonian of the model is the dimerized quantum $XY$ chain in the presence
of uniform ($h$) and alternating ($h_a$) transverse magnetic fields:
\begin{widetext}
\begin{equation}
\label{XYHam}
   H =\sum_{n=1}^{N}~  \frac{J}{4}  \Big[ (1+\gamma+\delta (-1)^{n} )
 \sigma_{n}^{x}\sigma_{n+1}^{x}
 + (1-\gamma+ \delta (-1)^{n}) \sigma_{n}^{y}\sigma_{n+1}^{y} \Big]
 + \frac12 \big[ h+(-1)^{n} h_a \big] \sigma_{n}^{z} ~.
\end{equation}
\end{widetext}
Here $\sigma$-s are the standard Pauli matrices, $J$ is the nearest-neighbor
exchange coupling, and $\gamma$ and
$\delta$ are the anisotropy and dimerization parameters,
respectively. This exactly-solvable model was first introduced and analyzed by
Perk \textit{et al}\cite{Perk75}. (See also \cite{DelGamMod,Lima,Sen2008,GT2017} for related more recent work on versions of this model.)
The standard Jordan-Wigner (JW) transformation \cite{Lieb61,Franchini2017} maps \eqref{XYHam} onto the free-fermionic
Hamiltonian
\begin{widetext}
\begin{equation}
\label{XYFermi}
   H =\sum_{n=1}^{N}~  \frac{J}{2}  \Big[ (1+\delta (-1)^{n} )(c_{n}^\dag c_{n+1} +\mathrm{h.c.})+
                                          \gamma (c_{n}^\dag c_{n+1}^\dag +\mathrm{h.c.}) \Big]
             +  \big( h+(-1)^{n} h_a \big) \big( c_{n}^\dag c_{n}- \frac12~ \big)  ~,
\end{equation}
\end{widetext}
called in recent literature the (modulated) Kitaev chain.\cite{Kitaev2001} In the fermionic representation \eqref{XYFermi} the
chain has dimerized hopping and modulated chemical potential. So, whether we deal with the modulated $XY$ spin or the
Kitaev fermionic chains, is a matter of mere convention, especially since the results below will be given in terms of spins or fermions,
on the same footing.

The duality transformation is defined as \cite{Perk,Fradkin78}
\begin{eqnarray}
  \sigma_{n}^{x} &=& \tau_{n-1}^{x}\tau_{n}^{x}
  \label{sigmaX} \\
  \sigma_{n}^{y} &=& \prod_{l=n}^{N} \tau_{l}^{z}~,
  \label{sigmaY}
\end{eqnarray}
where $\tau$ obey the standard algebra of the Pauli operators and reside on the sites of the dual lattice
which can be placed between the sites of the original chain.
This transformation maps \eqref{XYHam} onto the dual spin Hamiltonian
%\begin{widetext}
\begin{eqnarray}
  H= H_e + H_o + H_{mix} \label{Hsum} \\
  H_e= \frac{J}{4} \sum_{l=1}^{N/2}
    (1+\gamma -\delta) \tau_{2l-2}^{x} \tau_{2l}^{x}+ (1-\gamma +\delta)\tau_{2l}^{z}  \label{He} \\
  H_o= \frac{J}{4} \sum_{l=1}^{N/2}
    (1+\gamma +\delta) \tau_{2l-1}^{x} \tau_{2l+1}^{x}+ (1-\gamma -\delta)\tau_{2l-1}^{z}  \label{Ho} \\
 H_{mix} =-\frac{i}{2}\sum_{n=1}^{N} (h+ (-1)^n h_a)  \tau_{n-1}^{x} \tau_{n}^{x} \prod_{m=n}^{N} \tau_{m}^{z}~, \label{Hmix}
\end{eqnarray}
%\end{widetext}
which is a sum of two 1D transverse Ising models residing on the even and odd sites of the dual lattice, plus the field-induced
$\mathbb{Z}_2 \otimes \mathbb{Z}_2$ symmetry-breaking term $H_{mix}$ which couples the even and odd sectors of the dual
$\tau$-Hamiltonian.

%
%xxxxxxxxxxxxxxxxxxxxxxxxxxxxxxxxxxxxxxxxxxxxxxxxxxxxxxxxxxxxxxxxxxxxxxxxxxxxxx
%
\subsection{Spectrum}
%
%xxxxxxxxxxxxxxxxxxxxxxxxxxxxxxxxxxxxxxxxxxxxxxxxxxxxxxxxxxxxxxxxxxxxxxxxxxxxxx
%
The Hamiltonian \eqref{XYFermi} can also be
written as
\begin{equation}
\label{Hspinor}
 H= \frac12 \sum_{k}\psi^{\dag}_{k}\mathcal{H}(k) \psi_{k}~,
\end{equation}
where the fermions are unified in the spinor
\begin{equation}
  \psi_{k}^{\dag}=\left(c_1^{\dag}(k),
  c_2^{\dag}(k),c_1(-k), c_2(-k)\right)~,
\label{spinor1}
\end{equation}
with the wavenumbers restricted to the reduced Brillouin zone (BZ) $k \in
[-\pi/2,\pi/2]$ and we set the lattice spacing $a=1$. The band index $\alpha=1,2$ serves to map the JW fermions
from the full $2 \pi$-periodic BZ onto the reduced zone as
\begin{equation}
\label{c12}
  c(k)= c_1(k) \cdot \vartheta( \pi/2- |k|)+c_2(k-\pi) \cdot \vartheta(|k|- \pi/2)~,
\end{equation}
where $\vartheta(x)$ is the Heaviside step function. The JW fermion in the coordinate representation \eqref{XYFermi} is:
\begin{equation}
\label{cFourier}
  c_n=  \frac{1}{\sqrt{N}} \sum_{\alpha,q} c_\alpha(q) (-1)^{(\alpha-1)n} e^{-iqn} ~.
\end{equation}
The $4\times 4$ Hamiltonian  matrix (we set $J=1$ from now on) can be written as
\begin{equation}
\label{Hk}
  \mathcal{H}(k) = \left(%
\begin{array}{cc}
  \hat{A} & \hat{B} \\
  \hat{B}^\dag  & -\hat{A} \\
\end{array}%
\right)~,
\end{equation}
with
\begin{equation}
\label{A}
  \hat{A} \equiv  \left(%
\begin{array}{cc}
  h +\cos k & h_a+i \delta \sin k \\
  h_a-i \delta \sin k & h -\cos k \\
\end{array}%
\right)~,
\end{equation}
and
\begin{equation}
\label{B}
 \hat{B} \equiv  \left(%
\begin{array}{cc}
  -i \gamma \sin k & 0 \\
  0 & i \gamma \sin k \\
\end{array}%
\right)~,
\end{equation}
The Hamiltonian has four eigenvalues\cite{Perk75} $\pm E_{\pm}$ where
\begin{equation}
\label{Epm}
 E_{\pm}(k)=\sqrt{\mathfrak{C}_2(k)\pm \sqrt{\mathfrak{C}_2^2(k)-\mathfrak{C}_4(k)}}~,
\end{equation}
with
\begin{equation}
  \label{C2}
  \mathfrak{C}_2(k) \equiv h^2+h_a^2+\cos^2k+(\delta^2+\gamma^2)\sin^2k
\end{equation}
and
\begin{equation}
  \label{C4}
  \mathfrak{C}_4(k) \equiv \Big(h^2-h_a^2-\cos^2k-(\delta^2-\gamma^2)\sin^2k \Big)^2 +(\gamma \sin 2k)^2
\end{equation}

%
%xxxxxxxxxxxxxxxxxxxxxxxxxxxxxxxxxxxxxxxxxxxxxxxxxxxxxxxxxxxxxxxxxxxxxxxxxxxxxx
%
\subsection{Phase diagram}
\begin{figure}[h]
\centering{\includegraphics[width=8.5cm]{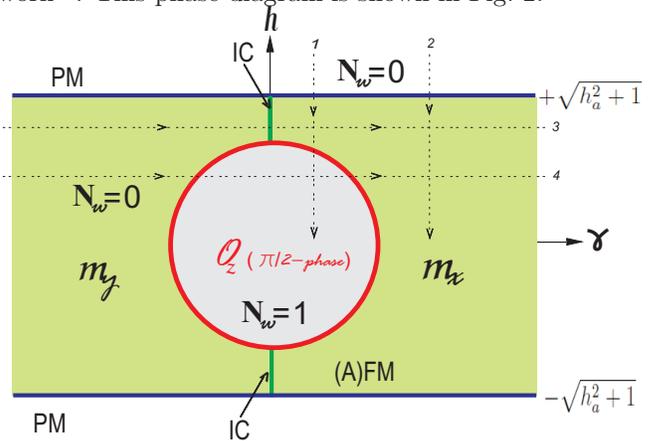}}
\caption{Phase diagram of the model in $h-\gamma$ plane. The model is critical  \textit{(i)} on two infinite
lines $|h|=\sqrt{h_a^2+1}$; \textit{(ii)} on the circle $h^2+\gamma^2 =h_a^2+\delta^2$;
\textit{(iii)} on two line segments $\sqrt{h_a^2+\delta^2} \leq |h| \leq \sqrt{h_a^2+1}$ at  $\gamma= 0$.
Depending on sign of the spin coupling $J$, the local order $m_{x,y}$ can be ferro- or antiferromagnetic.
Three phases: disordered paramagnetic (PM), (anti)ferromagnetic, and topological with modulated string order
parameter $\mathcal{O}_z$ are shown. The four paths (1-4) in parametric space used for calculation of magnetization
and string order parameters are indicated by thin dashed lines. The winding numbers $N_w$ calculated in Sec.~\ref{OPs} are also shown.}
\label{Hgamma}
\end{figure}

%
%xxxxxxxxxxxxxxxxxxxxxxxxxxxxxxxxxxxxxxxxxxxxxxxxxxxxxxxxxxxxxxxxxxxxxxxxxxxxxx
%
The phase diagram of the model was first found by Perk \textit{et al}\cite{Perk75}.
See also \cite{TIMbook} for a recent review.
The critical lines where the model becomes gapless are determined by the condition
$\mathfrak{C}_4(k)=0$. Cf. eqs. \eqref{Epm}, \eqref{C4} and Fig.~\ref{Hgamma}.
There are three phase boundaries:\\
\textit{(i)} at
\begin{equation}
 \label{Hc}
 h=\pm\sqrt{h_a^2+1}~,~ \forall~ \gamma,\delta
\end{equation}
the gap vanishes at the center of the BZ ($k=0$). \\
\textit{(ii)} At the edge of the BZ ($k= \pm \pi/2$) the gap vanishes on the circle
\begin{equation}
 \label{Circle}
  h^2+\gamma^2=h_a^2+\delta^2~.
\end{equation}
\textit{(iii)} Two critical line segments at $\gamma=0$ correspond to the gap vanishing at the incommensurate (IC)
wavevector
\begin{equation}
 \label{kIC}
  k_c= \pm \arcsin \sqrt{\frac{1+h_a^2-h^2}{1-\delta^2}}~.
\end{equation}
The IC solution exists in the range of parameters:
\begin{equation}
 \label{ICrange}
 \gamma=0, ~ |\delta| < 1~~\mathrm{and}~~\sqrt{h_a^2+\delta^2} \leq |h| \leq \sqrt{1+h_a^2}~.
\end{equation}
The wavevector \eqref{kIC} varies continuously from $k_c=0$ at the intersection of $\gamma=0$ and $h=\pm\sqrt{h_a^2+1}$
to $k_c= \pm \pi/2$ where the critical segments end at the intersections with the circle.

\begin{figure}[h]
\centering{\includegraphics[width=8.5cm]{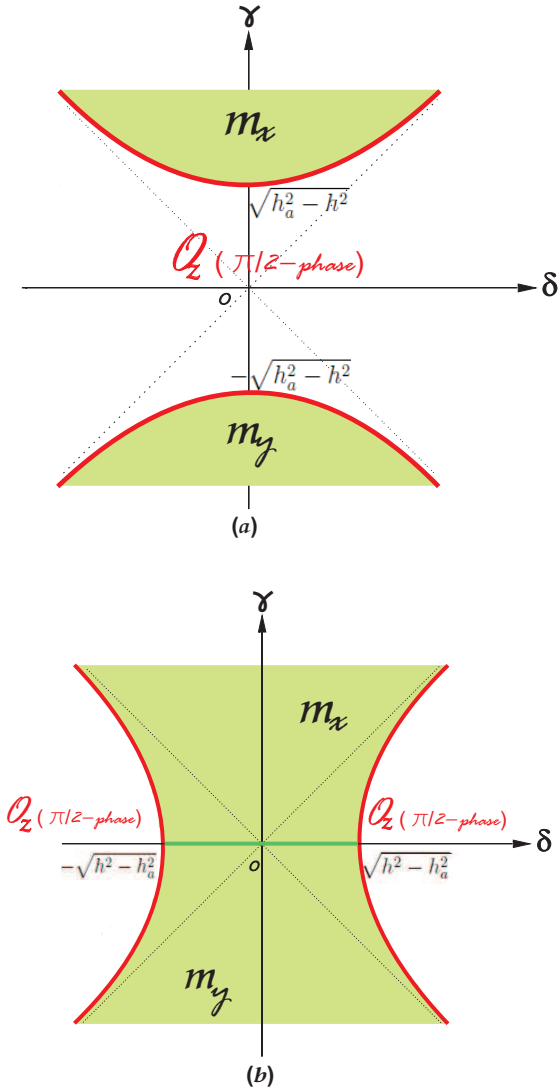}}
\caption{ The boundaries \eqref{Circle} between the phases with local order $m_{x,y}$ and with the $\pi/2$-modulated topological string order $\mathcal{O}_z$ in $\delta-\gamma$ plane for the cases $h<h_a$ and $h>h_a$. The two thin dotted lines $\gamma = \pm  \delta$ are
the phase boundaries in the limit $h=h_a=0$.\cite{GT2017} }
\label{DelGam}
\end{figure}

It is useful to plot the phase boundary \eqref{Circle} in the $\gamma-\delta$ plane, especially keeping in mind connection
to the earlier work \cite{GT2017}. This phase diagram is shown in Fig.~\ref{DelGam}.
%
%
%
%xxxxxxxxxxxxxxxxxxxxxxxxxxxxxxxxxxxxxxxxxxxxxxxxxxxxxxxxxxxxxxxxxxxxxxxxxxxxxx
%
\subsection{Field-induced magnetizations}
\begin{figure}[h]
\centering{\includegraphics[width=9.0cm]{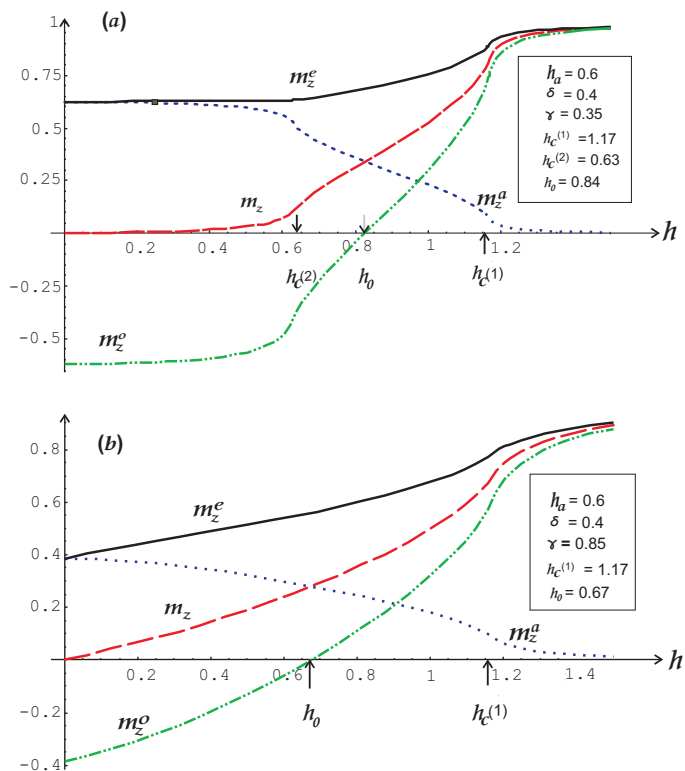}}
        \caption{Four magnetizations \textit{vs} uniform magnetic field $h$ at $\delta=0.4$ and fixed value of the alternated
        magnetic field $h_a=0.6$ for two values of $\gamma$.
        At the first panel (a) $\gamma=0.35$ and the field $h$ takes the path $1$ shown in the phase diagram
        Fig.~\ref{Hgamma}. It crosses the phase boundaries at $h=h_c^{(1)}=1.17$ and $h=h_c^{(2)}=0.63$ where the cusps are noticeable.
        The intermediate point $h_0=0.84$ where the odd sublattice magnetization vanishes and the induced magnetic pattern changes from
        ferrimagnetic to antiferrimagnetic (see Fig.~\ref{Spins}) is not related to any singularities.
        The order becomes antiferromagnetic at $h=0$.
        At the second panel (b) the path $2$ of the field $h$ (see Fig.~\ref{Hgamma}) does not cross the critical circle at $\gamma=0.85$.
        The magnetizations demonstrate cusps at the only critical point $h_c^{(1)}$, while at $h< h_c^{(1)}$, including the point
        $h=h_0=0.67$ of the induced pattern switch, the magnetizations are smooth.}
\label{Mzall}
\end{figure}

\begin{figure}[b]
\centering{\includegraphics[width=6cm]{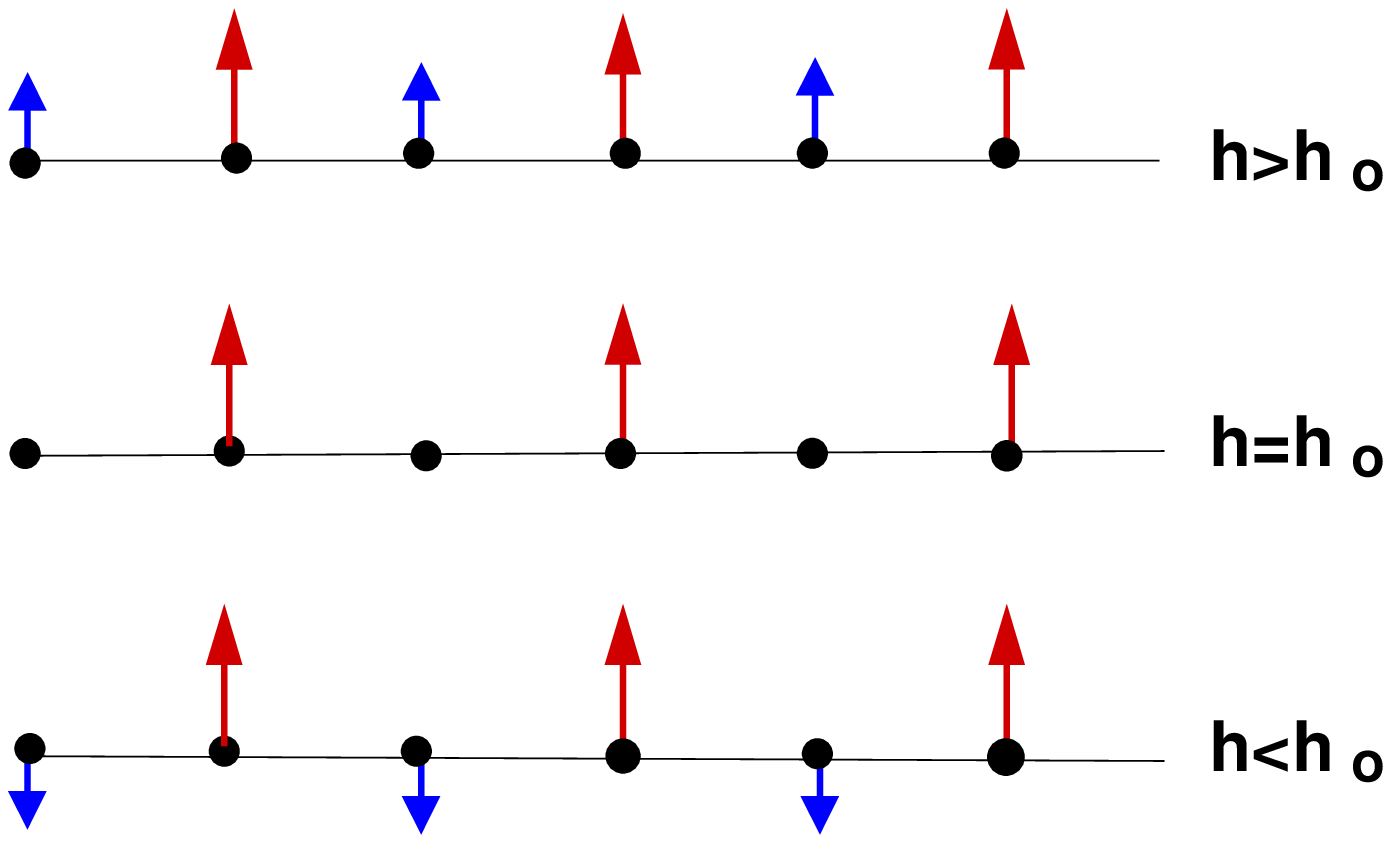}}
        \caption{Visualization of the field-induced magnetization at different values of uniform external field $h$ as presented
        in Fig.~\ref{Mzall}. The red/blue spins contribute to the even/odd transverse magnetizations, respectively.
        The pattern evolves smoothly with $h$ from ferrimagnetic ordering at $h>h_0$ to antiferrimagnetic
        at $h<h_0$ passing through the point $h=h_0$ where the magnetization on the odd sites $m_z^o$ vanishes.}
\label{Spins}
\end{figure}

%
%xxxxxxxxxxxxxxxxxxxxxxxxxxxxxxxxxxxxxxxxxxxxxxxxxxxxxxxxxxxxxxxxxxxxxxxxxxxxxx
%
%
Differentiation of the free energy with respect to $h$ and to $h_a$ yields magnetizations
\begin{equation}
\label{mz}
 m_z = \frac1N \sum_{n=1}^{N} \langle \sigma_{n}^{z} \rangle
\end{equation}
and
\begin{equation}
\label{mza}
 m_z^a = \frac1N \sum_{n=1}^{N}(-1)^n \langle \sigma_{n}^{z} \rangle~,
\end{equation}
respectively.\cite{Perk75}
At zero temperature the explicit expressions are:
\begin{widetext}
\begin{equation}
\label{mz1}
m_z= \frac{h}{\pi} \int_{0}^{\frac{\pi}{2}} \left\{\Big(\frac{1}{E_{+}}+\frac{1}{E_{-}}\Big)+
\frac{\cos^{2}k+|w_a|^2 }{\sqrt{h^2\cos^2 k+|w|^2|w_a|^2}}
\Big(\frac{1}{E_{+}}-\frac{1}{E_{-}}\Big)\right\}dk
\end{equation}
and
\begin{equation}
\label{mza1}
m_z^a= \frac{h_a}{\pi} \int_{0}^{\frac{\pi}{2}} \left\{ \Big(\frac{1}{E_{+}}+\frac{1}{E_{-}}\Big)+
\frac{ |w|^2}{\sqrt{h^2\cos^2 k+|w|^2|w_a|^2}}
\Big(\frac{1}{E_{+}}-\frac{1}{E_{-}}\Big)\right\}dk
\end{equation}
\end{widetext}
where we defined the auxiliary parameters:
\begin{equation}
\label{zza}
w \equiv h+i\gamma\sin k~~\mbox{and}~~ w_a \equiv h_a+i\delta\sin k~.
\end{equation}
The physics becomes more transparent if we introduce two magnetizations on even/odd sublattices $m_z^{e/o}$ as
\begin{equation}
\label{mzoe}
m_z=\frac12 ( m_z^e +m_z^o),~~m_z^a=\frac12 ( m_z^e -m_z^o) ~.
\end{equation}
We plot all four field-induced magnetizations as functions of the uniform component of the magnetic field $h$ in Fig.~\ref{Mzall}.
Two cases need to be distinguished. The first case shown in panel (a) corresponds to the variation of the field $h$ along
the path $1$ on the phase diagram in the $h-\gamma$ plane shown  in Fig.~\ref{Hgamma}. The path crosses the PM-FM phase boundary
at $h=h_c^{(1)}=\sqrt{1+h_a^2}$ and the boundary between the ferromagnetic and topological phases at
$h=h_c^{(2)}=\sqrt{h_a^2+\delta^2-\gamma^2}$. The magnetizations have noticeable cusps at these critical points, which after
differentiation result in divergent susceptibilities.\cite{Perk75} At the intermediate field $h_0$ (its value is available numerically
only) the odd sublattice magnetization vanishes and the induced magnetic pattern changes from ferrimagnetic to antiferrimagnetic
(see Fig.~\ref{Spins}). This point is not related to any singularities in magnetizations or their derivatives.

The second case shown in panel (b) corresponds to the path $2$ on the phase diagram. It crosses only the PM-FM phase boundary and bypasses the
topological phase. The magnetizations demonstrate cusps at the only critical point $h_c^{(1)}$, while at $h< h_c^{(1)}$, including the point
$h=h_0=0.6758$ of the induced pattern switch, the magnetizations and their derivatives are analytical.

The main conclusions following from the analysis of the field-induced magnetization are: \textit{(i)} the magnetization cusps which translate into corresponding diverging susceptibilities do probe the critical points (phase boundaries); \textit{(ii)} the vanishing odd sublattice magnetization
$m_z^o$ and related change of direction of the odd magnetization at the intermediate field $h=h_0$ do not constitute a critical point; \textit{(iii)}
the components of transverse magnetization cannot probe the order (or serve to build up a local order parameter) of the phase lying inside the circle shown in Fig.~\ref{Hgamma}.
%
%
%
%%%%%%%%%%%%%%%%%%%%%%%%%%%%%%%%%%%%%%%%%%%%%%%%%%%%%%%%%%%%%%%%%%%%%%%%%%%%%%
%
\section{Order Parameters}\label{OPs}
%
%%%%%%%%%%%%%%%%%%%%%%%%%%%%%%%%%%%%%%%%%%%%%%%%%%%%%%%%%%%%%%%%%%%%%%%%%%%%%%
%
%
%xxxxxxxxxxxxxxxxxxxxxxxxxxxxxxxxxxxxxxxxxxxxxxxxxxxxxxxxxxxxxxxxxxxxxxxxxxxxxx
%
\subsection{Bogoliubov tranformation and Majorana operators}
%
%xxxxxxxxxxxxxxxxxxxxxxxxxxxxxxxxxxxxxxxxxxxxxxxxxxxxxxxxxxxxxxxxxxxxxxxxxxxxxx
%
%
To diagonalize the Hamiltonian \eqref{Hspinor} in terms of the new fermionic operators
$(\eta_\alpha(q),\eta_\alpha^{\dag}(q))$, we utilize the Bogoliubov canonical transformation
within the formalism worked out in \cite{Lieb61,Lima}.
It is convenient to introduce the Majorana fermions as
\begin{equation}
\label{Maj}
   a_n +i b_n  \equiv 2 c^{\dag}_n~.
\end{equation}
Then the Bogoliubov transformation reads as
\begin{widetext}
\begin{eqnarray}
  a_n &=&  \frac{1}{\sqrt{N}} \sum_{\alpha,\beta,q} \Psi^\ast_{\beta \alpha}(q)
  \Big[ \eta_\beta(q) + \eta_\beta^{\dag}(-q) \Big] (-1)^{(\alpha-1)n} e^{-iqn} \label{an} \\
  i b_n &=&  \frac{-1}{\sqrt{N}} \sum_{\alpha,\beta,q} \Phi^\ast_{\beta \alpha}(q)
  \Big[ \eta_\beta(q) - \eta_\beta^{\dag}(-q) \Big] (-1)^{(\alpha-1)n} e^{-iqn} \label{bn}
\end{eqnarray}
\end{widetext}
The unitary $2 \times 2$ matrices $\hat \Phi$ and $\hat \Psi$
\begin{equation}
\label{Norm}
  \hat{\Phi}^\dag \hat{\Phi}=\hat{\Psi}^\dag \hat{\Psi}=\hat{\mathbb{1}}
\end{equation}
are constructed from the normalized (left) eigenvectors of the operators
$(\hat{A} \mp \hat{B})(\hat{A} \pm \hat{B})$ whose eigenvalues are $E^2_\pm$. Explicitly,  $\hat \Phi$ and $\hat \Psi$
solve the following equations:
\begin{eqnarray}
  \hat{\Phi}(\hat{A} - \hat{B})(\hat{A} + \hat{B}) &=& \hat{I}_E^2 \hat{\Phi} \label{PhiEq}\\
  \hat{\Psi}(\hat{A} + \hat{B})(\hat{A} - \hat{B}) &=& \hat{I}_E^2 \hat{\Psi}~, \label{PsiEq}
\end{eqnarray}
where
\begin{equation}
\label{Ie}
  \hat{I}_E \equiv \mathrm{diag}(E_+, E_-)~.
\end{equation}
In addition, these matrices satisfy the conditions:
\begin{eqnarray}
  \Phi^\ast_{\beta \alpha}(-q) &=& \Phi_{\beta \alpha}(q) \\
  \Psi^\ast_{\beta \alpha}(-q) &=& \Psi_{\beta \alpha}(q) \label{Reality}
\end{eqnarray}
We find
\begin{equation}
\label{Phi}
   \hat \Phi(q)= \left(
                   \begin{array}{cc}
                    e^{-i \theta} \beta_+ & \beta_- \\
                     -\beta_- & e^{i \theta} \beta_+ \\
                   \end{array}
                 \right)~,
\end{equation}
where
\begin{equation}
\label{theta}
  e^{i \theta}  \equiv \frac{w w_a}{|w| |w_a|} ~,
\end{equation}
and
\begin{equation}
\label{betapm}
  \beta_\pm \equiv \frac{1}{\sqrt{2}} \Big(  1 \pm \frac{h \cos q}{R} \Big)^\frac12~,
\end{equation}
with
\begin{equation}
\label{R}
  R \equiv \sqrt{h^2\cos^2 q+|w|^2|w_a|^2}
\end{equation}
One can also parameterize $\hat \Phi$ in terms of the Bogoliubov ange $\vartheta_B$ defined as
\begin{equation}
\label{thetaB}
  \beta_+ \equiv \cos \vartheta_B~,~~~~\beta_- \equiv \sin \vartheta_B~.
\end{equation}
Once the solution of \eqref{PhiEq} is found, the solution of \eqref{PsiEq} can be calculated straightforwardly as
\begin{equation}
\label{Psi}
  \hat \Psi =  \hat{I}_E^{-1} \hat \Phi (\hat{A} - \hat{B})~.
\end{equation}
(Alternatively, one can first find $\hat \Psi$  from \eqref{PsiEq}, and then $\hat \Phi$ as
$\hat \Phi =  \hat{I}_E^{-1} \hat \Psi (\hat{A} + \hat{B})$.)
Using the thermodynamic average for the Bogoliubov fermions
\begin{eqnarray}
  \langle  \eta_\alpha(q) \eta_\beta^{\dag}(q') \rangle &=& \delta_{\alpha \beta} \delta_{qq'} (1-n_\alpha (q))~, \\
  \langle  \eta_\alpha^\dag (q) \eta_\beta (q') \rangle &=& \delta_{\alpha \beta} \delta_{qq'} n_\alpha (q)~,
\end{eqnarray}
where $n_\alpha (q) =(1+\exp(\beta E_\alpha))^{-1} $ is the Fermi-Dirac distribution function, we can obtain the field-induced magnetization
\begin{equation}
\label{mzMatrix}
  m_z= \frac{1}{2 \pi} \int_{-\pi/2}^{\pi/2}
  dq \mathrm{Tr} \big\{ \hat{\Phi}^\dag (q) \hat{\Psi} (q)\big\}~, ~T=0~.
\end{equation}
The above formula is equivalent to Eq.~\eqref{mz1} obtained from differentiation of the partition function.
Introducing the matrix
\begin{equation}
\label{G}
  \hat{G} (q)  \equiv  \hat{\Psi}^\dag (q) \hat{\Phi} (q)
\end{equation}
we find the zero-temperature correlation function of the Majorana operators:
\begin{widetext}
\begin{equation}
\label{ab}
  \langle ib_n a_m \rangle = \frac{1}{2 \pi} \sum_{\alpha, \beta}   \int_{-\pi/2}^{\pi/2} dq
  G_{\beta \alpha}(q)(-1)^{(\alpha-1)n} (-1)^{(\beta-1)m}  e^{-iq(m-n)}~.
\end{equation}
%\end{widetext}
The explicit form of matrix $\hat G$ \eqref{G} is calculated from Eqs. \eqref{Phi}, \eqref{Psi}.
One can check that its components satisfy the following relations:
\begin{eqnarray}
  \label{Gsym}
  G_{21}(q \pm \pi) &=& G_{12}(q)~, \\
  G_{22}(q \pm \pi) &=& G_{11}(q)~.
\end{eqnarray}
Then Eq. \eqref{ab} can be simplified into
%\begin{widetext}
\begin{equation}
\label{abS}
  \langle ib_n a_m \rangle = \frac{1}{2 \pi}    \int_{-\pi}^{\pi}  dq e^{-iq(m-n)}
  \Big\{ G_{11}(q)+ (-1)^n G_{12}(q)  \Big\}~,
\end{equation}
where the matrix elements are found to be:
\begin{eqnarray}
  G_{11}(q) &=& (\cos q +w^\ast) \Big\{ \frac{\beta_+^2}{E_+}+ \frac{\beta_-^2}{E_-} \Big\} +
   \frac12 \frac{w^\ast |w_a|^2}{R}  \Big\{ \frac{1}{E_+}- \frac{1}{E_-} \Big\}~,  \label{G11} \\
    G_{12}(q) &=& w_a \Big\{ \frac{\beta_-^2}{E_+}+ \frac{\beta_+^2}{E_-} \Big\}+
    (\cos q +w^\ast) e^{i \theta} \beta_+ \beta_-
     \Big\{ \frac{1}{E_+}- \frac{1}{E_-} \Big\}~.  \label{G12}
\end{eqnarray}
\end{widetext}
The formulas derived in this subsection provide us with the main results needed for the rest of calculations.
%
%
%xxxxxxxxxxxxxxxxxxxxxxxxxxxxxxxxxxxxxxxxxxxxxxxxxxxxxxxxxxxxxxxxxxxxxxxxxxxxxx
%
\subsection{Spontaneous magnetization}
%
%xxxxxxxxxxxxxxxxxxxxxxxxxxxxxxxxxxxxxxxxxxxxxxxxxxxxxxxxxxxxxxxxxxxxxxxxxxxxxx
%
%
We define the spontaneous longitudinal sublattice magnetizations as
\begin{eqnarray}
  \langle \sigma_{2l}^x \sigma_{2m}^x \rangle &\xrightarrow[(m-l) \to \infty]{~}& (m_x^e)^2~, \label{mxe} \\
  \langle \sigma_{2l-1}^x \sigma_{2m-1}^x \rangle &\xrightarrow[(m-l) \to \infty]{~}& (m_x^o)^2~. \label{mxo}
\end{eqnarray}
The spontaneous longitudinal magnetization is the (local) order parameter defined as
\begin{equation}
\label{mx}
  m_x = \frac12 (m_x^e+m_x^o)
\end{equation}
We also define the Majorana string operator:
\begin{equation}
\label{Ox}
  O_x(m) = \prod_{l=1}^{m-1} \big[ i b_l a_{l+1} \big]~.
\end{equation}
(By definition $O_x(1)=1$.)
For further reference let us remind some useful relations between original spins, Majorana fermions, and the dual spin operators
(\ref{sigmaX},\ref{sigmaY}):
\begin{eqnarray}
  \sigma_{n}^{x}  \sigma_{n+1}^{x} &=& i b_n a_{n+1}= \tau_{n-1}^{x}\tau_{n+1}^{x}
  \label{XX} \\
  \sigma_{n}^{y} \sigma_{n+1}^{y} &=& -i a_n b_{n+1} =\tau_{n}^{z}~.
  \label{YY}
\end{eqnarray}
Then $O_x(m)=\sigma_{1}^{x} \sigma_{m}^{x}$.
The spin-correlation function can be calculated as the
correlation function of the Majorana string operators: \cite{Lieb61}
\begin{equation}
\label{OxMx}
  \langle \sigma_L^x \sigma_R^x \rangle = \langle O_x(L) O_x(R)  \rangle =
  \langle  \prod_{n=L}^{R-1} \big[ i b_n a_{n+1} \big] \rangle~.
\end{equation}
The latter is given by the determinant:
\begin{widetext}
\begin{equation}
\label{mxDet}
  \langle \sigma_L^x \sigma_R^x \rangle =
  \left|
    \begin{array}{cccc}
      \langle ib_L a_{L+1}\rangle & \langle ib_L a_{L+2}\rangle & \ldots & \langle ib_L a_{R}\rangle \\[0.2cm]
      \langle ib_{L+1} a_{L+1}\rangle & \langle ib_{L+1} a_{L+2}\rangle & \ldots &  \langle ib_{L+1} a_{R}\rangle \\[0.25cm]
      \vdots & \ddots & \ddots & \vdots \\[0.25cm]
      \langle ib_{R-1} a_{L+1}\rangle & \langle ib_{R-1} a_{L+2}\rangle & \ldots &  \langle ib_{R-1} a_{R}\rangle \\
    \end{array}
  \right|
\end{equation}
\end{widetext}
To calculate the quantities of our interest we choose the ends as:
\begin{eqnarray}
  L=2,~R=2N ~ &\longmapsto&~ \mathrm{even~quantity}~,\nonumber \\
   L=1,~R=2N-1 ~ &\longmapsto&~ \mathrm{odd~quantity} ~.
  \label{LMpm}
\end{eqnarray}
In both cases we are dealing with $2(N-1) \times 2(N-1)$ matrices. Note that \eqref{mxDet} is not the Toeplitz determinant, since the elements of
the matrix  $\langle ib_n a_m \rangle$ given by Eq.\eqref{abS} do not satisfy the condition $\langle ib_n a_m \rangle =f(m-n)$. It is however possible
to represent \eqref{mxDet} via the block Toeplitz matix. \cite{Widom70,Basor2019} Let us define
\begin{equation}
 \label{Gpm}
  G^\pm (q) \equiv  G_{11}(q) \pm G_{12}(q)
\end{equation}
and its inverse Fourier transform
\begin{equation}
 \label{GpmX}
  G^\pm (l) =\frac{1}{2 \pi}    \int_{-\pi}^{\pi}  dq e^{-iql}  G^\pm (q)~,
\end{equation}
along with the $2 \times 2$ matrix
\begin{equation}
\label{GlOp}
  \hat{\mathcal{G}}_{e/o} (l) \equiv
  \left(
    \begin{array}{cc}
      G^\pm (l)  &  G^\pm (l+1) \\
      G^\mp (l-1) &  G^\mp (l) \\
    \end{array}
  \right)~.
\end{equation}
Then the sublattice magnetizations $m_x^{e/o}$ can be evaluated as the limit of the determinant of the corresponding block Toeplitz matrix constructed from $(N-1)\times(N-1)$ blocks  $\hat{\mathcal{G}}_{e/o}$  of size $2 \times 2$:
\begin{widetext}
\begin{equation}
\label{mxBlock}
  \left|
    \begin{array}{cccc}
     \hat{\mathcal{G}}_\sharp (-1) & \hat{\mathcal{G}}_\sharp (1) & \ldots & \hat{\mathcal{G}}_\sharp (3-2N) \\[0.2cm]
      \hat{\mathcal{G}}_\sharp (1) & \hat{\mathcal{G}}_\sharp (-1) & \ldots &  \hat{\mathcal{G}}_\sharp (5-2N) \\[0.25cm]
      \vdots & \ddots & \ddots & \vdots \\[0.25cm]
      \hat{\mathcal{G}}_\sharp (2N-5) & \hat{\mathcal{G}}_\sharp (2N-7) & \ldots &  \hat{\mathcal{G}}_\sharp (-1) \\
    \end{array}
  \right| \xrightarrow[N \to \infty]{~} (m_x^\sharp )^2~.
\end{equation}
\end{widetext}
At this point we were unable to derive analytical results for asymptotics of the above block Toeplitz determinants. So we resort
to direct numerical calculations for large finite-size matrices. The results for spontaneous magnetization are given in
Fig.~\ref{OPsall}. The numerical values of the parameters we present in that figure are stable in the fourth decimal place for the
$M \times M$ matrices of  sizes $M \gtrsim 30$. In immediate vicinities of the critical points the order parameters are checked to decay smoothly as
$M \to \infty$.

We checked that the numerical results obtained from  \eqref{mxBlock} agree with two available analytical limits at $h_a=0$.
\begin{equation}
\label{ha0}
  h_a=0~\longmapsto~m_x^e =m_x^o =m_x~.
\end{equation}
For $\delta =0$ the result is due to Pfeuty: \cite{Pfeuty70}
\begin{equation}
 \label{hg}
  m^2_{x}=\frac{2}{1+\gamma}\left[\gamma^2(1-h^2)\right]^{1/4}, ~h<1~.
\end{equation}
For the $h =0$  case the order parameter can be obtained by combining the result of Pfeuty and the duality
transformations \eqref{sigmaX} and \eqref{sigmaY}, yielding \cite{GT2017}
\begin{equation}
 \label{dg}
 m^2_{x}=2\left[\frac{(\gamma^2-\delta^2)}{((1+\gamma)^2-\delta^2)^2} \right]^{1/4}, ~\delta < \gamma~.
\end{equation}
The analytical results \eqref{hg} and \eqref{dg} can be obtained from  \eqref{mxBlock} by the brut force calculations
utilizing Szeg\"{o}'s theorem \cite{McCoyBook}, but those calculations are quite demanding,
see Appendices for more details. 

The expressions for $m_y$ are obtained along the same lines. Numerical values satisfy
useful relation $m_y(-\gamma)=m_x(\gamma)$, see  Fig.~\ref{OPsall}.

%
%
%xxxxxxxxxxxxxxxxxxxxxxxxxxxxxxxxxxxxxxxxxxxxxxxxxxxxxxxxxxxxxxxxxxxxxxxxxxxxxx
%
\subsection{Nonlocal string order}
%
%xxxxxxxxxxxxxxxxxxxxxxxxxxxxxxxxxxxxxxxxxxxxxxxxxxxxxxxxxxxxxxxxxxxxxxxxxxxxxx
%
%
\begin{widetext}
We define another string operator
\begin{equation}
\label{Oz}
  O_z(n) \equiv \prod_{l=1}^{n} \big[ i b_{l} a_{l} \big]=\prod_{l=1}^{n} \sigma_l^z~,
\end{equation}
and the related string correlation function:
\begin{equation}
\label{CorrOz}
\langle  \prod_{l=L}^{R} \big[ i b_l a_l \big]  \rangle =
 \left|
    \begin{array}{cccc}
      \langle ib_L a_{L}\rangle & \langle ib_L a_{L+1}\rangle & \ldots & \langle ib_L a_{R}\rangle \\[0.2cm]
      \langle ib_{L+1} a_{L}\rangle & \langle ib_{L+1} a_{L+1}\rangle & \ldots &  \langle ib_{L+1} a_{R}\rangle \\[0.25cm]
      \vdots & \ddots & \ddots & \vdots \\[0.25cm]
      \langle ib_{R} a_{L}\rangle & \langle ib_{R} a_{L+1}\rangle & \ldots &  \langle ib_{R} a_{R}\rangle \\
    \end{array}
  \right|
\end{equation}
Inside the circle $|h| <h_c^{(2)}$ this string correlation function is found to be
oscillating with the period of four lattice spacing, see Fig.~\ref{ModSOP}, so we will label it as $\pi /2 $-phase to distinguish it from the
positive ``ferrimagnetic-like" string correlation function in the paramagnetic phase $|h| >h_c^{(1)}$.
\begin{figure}[h]
\centering{\includegraphics[width=7.5cm]{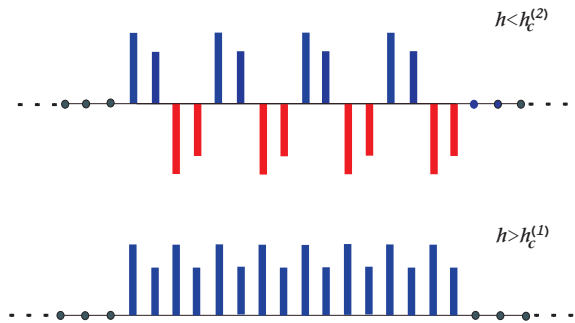}}
\caption{Visualization of the modulation of the string order parameter $\mathcal{O}_z$ inside the circle ($|h| <h_c^{(2)}$) on the phase diagram shown in Fig.~\ref{Hgamma} vs ``plain" behavior of $\mathcal{O}_z$  in the paramagnetic phase  $|h| >h_c^{(1)}$.}
\label{ModSOP}
\end{figure}

We will need three parameters to account for the string order:
\begin{equation}
\label{Ozeo}
 \langle  \prod_{l=L}^{R} \big[ i b_l a_l \big]  \rangle
 \xrightarrow[R \to \infty]{~}
  \left\{
    \begin{array}{c}
      (-1)^m \mathcal{O}_{z,1}^2~,~~~~~ L=1,~R=2m \\[0.2cm]
      (-1)^m \mathcal{O}_{z,2}^2~,~~~~ L=2,~R=2m \\[0.2cm]
      (-1)^{m+L} \mathcal{O}_{z,3}^2~,~~~~~ L=1,~R=2m+1~ \mathrm{or}~ L=2,~R=2m+1  \\
    \end{array}
   \right.
\end{equation}
The reason for this is  that due to the dimerization $\delta$ and the staggered field $h_a$, the value of the string correlation function depends not
only on the length of the string, but also on whether its ends $L/R$ are sitting on the even or odd sites of the chain.
The explicit matrix form \eqref{CorrOz} with the elements $G^\pm(i-j)$ can be written as the block Toeplitz matrix, similar to \eqref{mxBlock}.
Four-site periodic oscillations of the string correlation function inside the circle indicate periodicity of orientations of $\sigma_z$ spins along the string, or, alternatively, the modulation of fermionic density around half-filling.  In the FM phase (see Fig.~\ref{Hgamma}) the $O_z$ string
correlation function vanishes at large length, while in the paramagnetic phase $|h| >h_c^{(1)}$ it is positive, showing only quite trivial ``ferrimagnetic" oscillations synchronized with the staggered field. Qualitatively, it indicates that all $\sigma_z$ spins in the string are polarized along the field, or, in terms or fermions, the latter have concentration above half-filling.

In the limit  $h=h_a=0$ the hyperbolic phase boundaries shown in Fig.~\ref{DelGam} reduce to two lines $\gamma = \pm \delta$, \cite{GT2017}
and nonvanishing SOPs  $\mathcal{O}_{z,\sharp}$ are localized inside the cone $\delta^2>\gamma^2$.
When $\delta>0$ the only surviving component of the 4-periodic string correlation function is $\mathcal{O}_{z,1}$, such that
\begin{equation}
\label{OzDG+}
 \langle  \prod_{l=1}^{2m} \big[ i b_l a_l \big]  \rangle
 \xrightarrow[m \to \infty]{~}
 (-1)^m \mathcal{O}_{z,1}^2~,
\end{equation}
while $\mathcal{O}_{z,2}=\mathcal{O}_{z,3}=0$. Similarly, when $\delta<0$
\begin{equation}
\label{OzDG-}
 \langle  \prod_{l=2}^{2m} \big[ i b_l a_l \big]  \rangle
 \xrightarrow[m \to \infty]{~}
 (-1)^m \mathcal{O}_{z,2}^2~,
\end{equation}
while $\mathcal{O}_{z,1}=\mathcal{O}_{z,3}=0$. Note useful relations:
\begin{eqnarray}
  \label{ProdOzeo}
  O_{x,e}(m) O_{y,o}(m) &= & (-1)^{m} O_z(2m)~,  \\
  O_{x,o}(m) O_{y,e}(m) &= & (-1)^{m} O_z(1) O_z(2m+1) ~.
\end{eqnarray}
In the limit  $h=h_a=0$ the averaging of the even and odd strings decouples, and the SOPs $\mathcal{O}_{z,i}$ can be found in a simple form.
The values $\mathcal{O}_{x/y,e/o}$ inside the cone $\delta^2>\gamma^2$ are available \cite{GT2017} yielding
\begin{equation}
\label{OzAn}
  \left(
    \begin{array}{c}
      \mathcal{O}_{z,1} \\
      \mathcal{O}_{z,2} \\
    \end{array}
  \right) =
  \sqrt{2} \left[\frac{(\delta^2- \gamma^2)}{((1 \pm \delta)^2-\gamma^2)^2} \right]^{1/8} \cdot
   \left(
    \begin{array}{c}
     \vartheta(\delta) \\
      \vartheta(-\delta) \\
    \end{array}
   \right)~.
\end{equation}
We have checked the agreement between the analytical result \eqref{OzAn} and the numerical evaluation of the determinant \eqref{CorrOz}.

\begin{figure}[h]
\centering{\includegraphics[width=18cm]{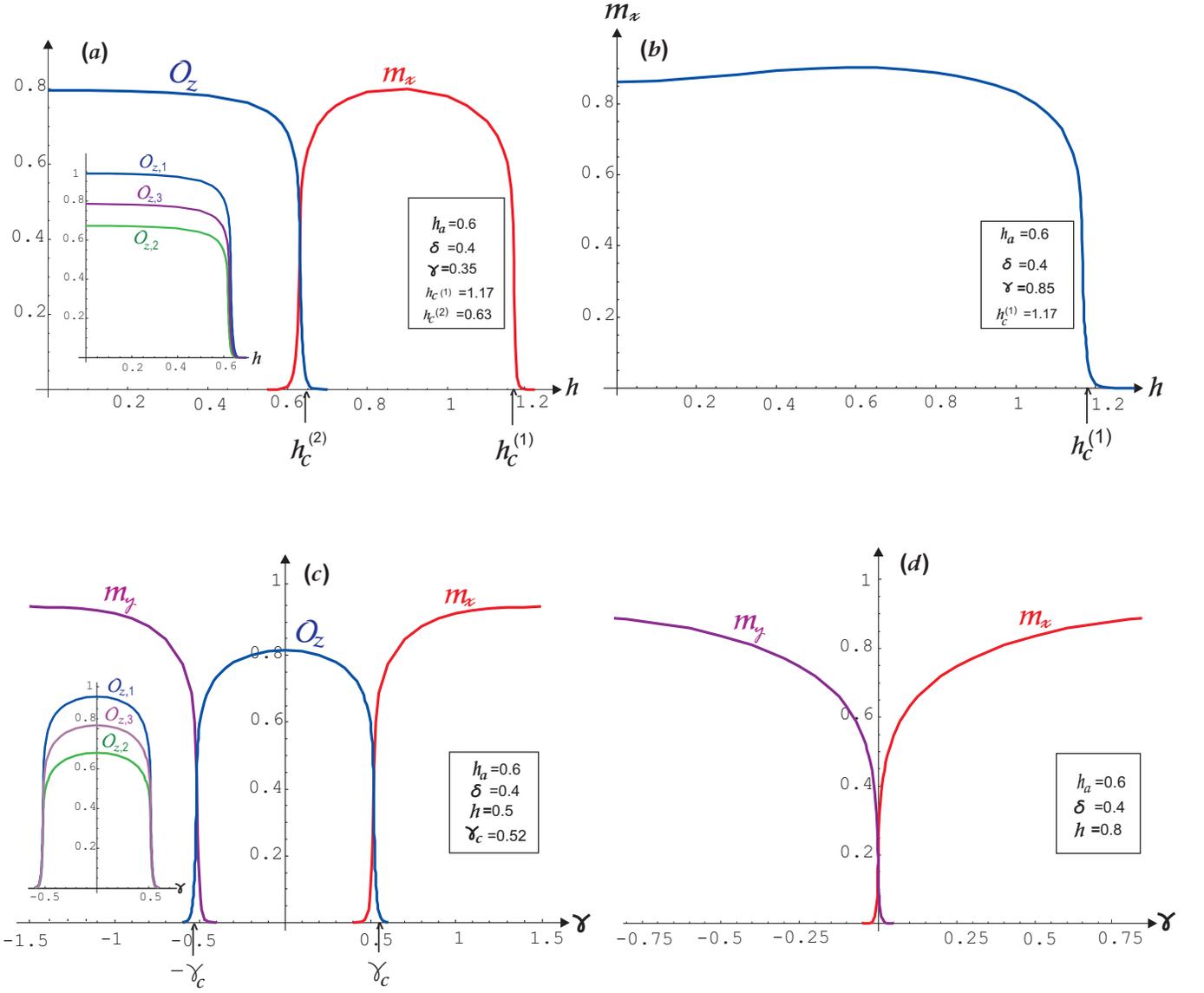}}
\caption{Spontaneous magnetizations $m_{x,y}$ and modulated string order parameter $\mathcal{O}_{z}$ numerically calculated from the $2N \times 2N$ matrices with $N=70$. The panels (a-d) correspond to the paths $1-4$
on the phase diagram shown in Fig.~\ref{Hgamma}. The spontaneous magnetizations are the averaged values of the even /odd components \eqref{mx} (not shown), while the modulated string order parameter $\mathcal{O}_{z}$ is the average value of the three parameters $\mathcal{O}_{z,i}$, shown in insets.
In the phases with local orders $m_{x,y}$ the string parameters $\mathcal{O}_{z,i}$ vanish, while in the disordered phase $h>h_c^{(1)}$ they are nonzero, but physically trivial (not shown).}
\label{OPsall}
\end{figure}
\end{widetext}

%
%
%xxxxxxxxxxxxxxxxxxxxxxxxxxxxxxxxxxxxxxxxxxxxxxxxxxxxxxxxxxxxxxxxxxxxxxxxxxxxxx
%
\subsection{Winding Number}\label{Ntop}
%
%xxxxxxxxxxxxxxxxxxxxxxxxxxxxxxxxxxxxxxxxxxxxxxxxxxxxxxxxxxxxxxxxxxxxxxxxxxxxxx
%
%
It has been shown in recent years that many quantum phase transitions with
hidden orders are accompanied by a change of topological numbers
\cite{FradkinBook13,TI}. Here we calculate the
winding number (or the Pontryagin index) in all regions of the model's phase diagram. Such parameters
were calculated recently in similar 1D models, see, e.g.,
\cite{Wu12,Niu12,EzawaNagaosa14,Zeng16,GT2017,Ezawa17,Miao17b}

By a unitary transformation the Hamiltonian \eqref{Hk} can be brought to the block off-diagonal form
\begin{equation}
\label{HD}
  \tilde{\mathcal{H}}(k)=\left(%
\begin{array}{cc}
  0 & \hat{D}(k) \\
  \hat{D}^\dag (k) & 0 \\
\end{array}%
\right)~,
\end{equation}
with the operator
\begin{equation}
\label{Dk}
    \hat{D}(k) \equiv \hat{A} (k)+\hat{B}(k)~,
\end{equation}
which has two eigenvalues
\begin{widetext}
\begin{equation}
\label{lambdapm}
    \lambda_\pm (k)=h \pm  \frac{1}{\sqrt{2}}
    \Big(1+2 h_a^2+\delta^2-\gamma^2+(1-\delta^2+\gamma^2) \cos 2k -2 i \gamma \sin 2k \Big)^{1/2}~.
\end{equation}
\end{widetext}
In one spatial dimension the winding number defined as \cite{SchnyderRyu11}
\begin{equation}
\label{Nw}
  N_w=\frac{1}{2\pi i} \int_{BZ} dk \mathrm{Tr}[\partial_k \ln \hat{D}]
\end{equation}
can be readily calculated for this model as:
\begin{equation}
\label{NwAn}
  N_w=\frac{1}{2\pi i} \arg(\lambda_+(k)+\lambda_-(k)) \Big|_{-\frac{\pi}{2}^+}^{\frac{\pi}{2}^-}~.
\end{equation}
The results for $N_w$ are given on the ground-state phase diagram in Fig.~\ref{Hgamma}.
$N_w$ can be viewed as a complimentary parameter characterizing a given phase.
The disordered (PM) phase and the magnetic phases where conventional local order $m_{x,y}$ exists, are topologically trivial,
$N_w=0$. The phase inside the circle where modulated nonlocal string order parameter $\mathcal{O}_z$ exists, is
topologically nontrivial, $N_w=1$.

%
%
%
%xxxxxxxxxxxxxxxxxxxxxxxxxxxxxxxxxxxxxxxxxxxxxxxxxxxxxxxxxxxxxxxxxxxxxxxxxxxxxx
%
\section{Conclusion}\label{Concl}
%
%xxxxxxxxxxxxxxxxxxxxxxxxxxxxxxxxxxxxxxxxxxxxxxxxxxxxxxxxxxxxxxxxxxxxxxxxxxxxxx
%

The main motivation for this work was to further advance the framework incorporating nonlocal
string order into an ``extended" Landau paradigm. For a large class of quantum spin or fermionic problems
we are interested in, the effective Ginzburg-Landau Hamiltonian to deal with, is a quadratic fermionic
Hamiltonian. In general such a Hamiltonian is already a result of some mean-field approximation, \cite{UsLadd,GT2017}
but there is a considerable number of physically interesting problems where it is the microscopic Hamiltonian of the model.
Postponing for future work building up the very important element --  the Wilsonian renormalization group appproach to
systematically deal with the nonlocal order beyond the mean field, we chose a non-interacting fermionic model to
analyze.

The model is the dimerized Kitaev chain with modulated chemical potential, which was initially introduced and studied
\cite{Perk75} as the dimerized $XY$ spin chain in the uniform and staggered transverse fields. These are two equivalent
representations of the model, since they map onto each other via the Jordan-Wigner transformation. This relatively simple
model is very relevant for studies of  quantum critical and out-of-equilibrium properties. \cite{TIMbook} The model has a rich
phase diagram (see Fig.~\ref{Hgamma}) which contains phases with local magnetic and nonlocal modulated string orders.

We have calculated the sponataneous magnetizations $m_{x,y}$ (local order parameters) showing that they smoothly vanish
at the corresponding phase boundaries via second-order quantum phase transitions. (Despite the fact that the model was
studied before, \cite{Perk75,Lima,Sen2008,TIMbook} we could not find the explicit results for magnetization in the previous
literature.) For the first time we have established the nature of the order in the topological phase lying inside the circle
in Fig.~\ref{Hgamma}. In that phase the modulated string order appears via a second-order phase transition. The modulations
are signalled by the oscillations of the $O_z$ string-string correlation function with the wave number $q=\pi/2$. Physically,
this correlation function probes the average of the string made out of $\sigma^z$ spins, or, equivalently, the string of
fermionic density operators (with respect to half-filling). In addition, we have calculated the winding number $N_w$ in all phases.
The disordered (PM) phase and the A(FM) phase with conventional local order parameter are topologically trivial, $N_w=0$,
while the phase with the modulated string order has $N_w=1$.  Form the results for the gaps and the free-fermionic nature
of the model we infer its critical indices to belong to the 2D Ising universality class. \cite{Note}

We need to stress once again \cite{GT2017,GYC2018} that there is no insurmountable difference between the local and string order parameters.
Using judiciously chosen duality transformations we show that a SOP can be identified as a local order parameter of some dual Hamiltonian.
Sometimes this can help to easily calculate the SOP in the dual framework, \cite{GT2017,GYC2018}, sometimes not. But it is important to understand as a matter of principle. For the general case when all model's parameters are nonzero  we were able to define the duality transformations reducing the SOPs to local ``dual" orders, but it did not result in technical simplifications in the calculations of SOPs.

On the technical side, the framework we present is quite straightforward: One needs to solve the problem of the Bogoliubov transformation which
allows to find explicit expression for the two-point Majorana correlation functions. The latter are building blocks of the Toepliz matrices.
The local and string order parameters, regardless of the original spin or fermionic representations, are given by the asymptotes of
the corresponding Majorana string correlation functions. The calculations of local and nonlocal parameters are reduced to the well-defined mathematical
problem of the evaluation of limits of determinants of the (block) Toeplitz matrices. These matrices are found in a closed form in terms
of the two-point Majorana correlation functions. With some luck and skills these limits can be found explicitly, \cite{McCoyBook} then one gets algebraic expressions for the order parameters. In this paper we found several expressions for the order parameters for particular limits of the model's couplings.
For the general case we were unable to do so.  The generalization of Szeg\"{o}'s theorem for the block Toeplitz matrices appeared to be a quite challenging mathematical problem. \cite{Widom70,Basor2019} This is however a simple numerical calculation, \cite{NoteMath} and our numerical results
are summarized in Fig.~\ref{OPsall}.

A very promising development of our results would be to find realizations of this model to experimentally
detect the predicted modulated string order. Very interesting questions of the IC gapless phase, disorder lines and the Majorana edge states in this
model will be addressed in a separate work.

%
%xxxxxxxxxxxxxxxxxxxxxxxxxxxxxxxxxxxxxxxxxxxxxxxxxxxxxxxxxxxxxxxxxxxxxxxxxxxxxx
\begin{acknowledgments}
G.Y.C. thanks the Centre for Physics of Materials at McGill University, where this
work was initiated, for hospitality. We are grateful to J.H.H. Perk for bringing
important papers to our attention and to Y.Y. Tarasevich for helpful comments.
Financial support from the Laurentian University Research
Fund (LURF), the Ministry of Education and Science of the Russian Federation
(state assignment grant No. 3.5710.2017/8.9), is gratefully acknowledged.
\end{acknowledgments}
%xxxxxxxxxxxxxxxxxxxxxxxxxxxxxxxxxxxxxxxxxxxxxxxxxxxxxxxxxxxxxxxxxxxxxxxxxxxxxx
%

%%%%%%%%%%%%%%%%%%%%%%%%%%%%%%%%%%%%%%%%%%%%%%%%%%%%%%%%%%%%%%%%%%%%%%%%%%%%%%
\begin{appendix}
\section{Separation of the Majorana Hamiltonian.}\label{AppA}
%%%%%%%%%%%%%%%%%%%%%%%%%%%%%%%%%%%%%%%%%%%%%%%%%%%%%%%%%%%%%%%%%%%%%%%%%%%%%%
%
%
In the main text of the paper we kept definitions and transformations consistent with those
of earlier related work \cite{GT2017,GYC2018} to preserve continuity in the series. In this
Appendix we will use some modified transformations which make the analysis of separability
of the Hamiltonian, analytical treatment of its limiting cases, and the symmetry, more transparent.
To this end we introduce two new species of Majorana fermions (compare to \eqref{Maj}) as
\begin{equation}
\label{Maj-ti}
   \tilde a_n +i \tilde b_{n-1}  \equiv 2 c^{\dag}_n~.
\end{equation}
Then the JW and duality transformations (\ref{sigmaX},\ref{sigmaY})
read: (compare to \eqref{XX} and \eqref{YY})
\begin{eqnarray}
  \sigma_{n}^{x}  \sigma_{n+1}^{x} &=& i \tilde b_{n-1} \tilde a_{n+1}= \tau_{n-1}^{x}\tau_{n+1}^{x}
  \label{XX-ti} \\
  \sigma_{n}^{y} \sigma_{n+1}^{y} &=& i \tilde b_n  \tilde a_n =\tau_{n}^{z}~.
  \label{YY-ti}
\end{eqnarray}

This transformation maps the original Hamiltonian \eqref{XYHam} (cf. also \eqref{Hsum}-\eqref{Hmix})
onto
\begin{widetext}
\begin{eqnarray}
  H&=& H_e + H_o + H_{mix} \label{Hsum-ti} \\
  H_e&=& \frac{iJ}{4} \sum_{l=1}^{N/2}
    (1+\gamma -\delta) \tilde b_{2l-2} \tilde a_{2l}+ (1-\gamma +\delta) \tilde b_{2l} \tilde a_{2l}  \label{He-ti} \\
  H_o&=& \frac{iJ}{4} \sum_{l=1}^{N/2}
     (1+\gamma +\delta) \tilde b_{2l-1} \tilde a_{2l+1}+ (1-\gamma -\delta) \tilde b_{2l-1} \tilde a_{2l-1} \label{Ho-ti} \\
 H_{mix} &=& -\frac{i}{2}\sum_{n=1}^{N} (h+ (-1)^n h_a)  \tilde b_{n-1} \tilde a_{n}~, \label{Hmix-ti}
\end{eqnarray}
\end{widetext}
Recombining two Majorana fermions into a (new) single JW fermion as
\begin{equation}
\label{c-ti}
  2 \tilde c^{\dag}_n   \equiv \tilde a_n +i \tilde b_{n}  ~.
\end{equation}
and Fourier-transforming it according to $\tilde c_n \mapsto  \tilde c_{e,o}(k)$ (with $e$ or $o$  for  $n=2l$ or $2l-1$, resp.),
we can bring the Hamiltonian into the spinor form \eqref{Hspinor} with
\begin{equation}
  \psi_{k}^{\dag}=\left(c_e^\dag(k), c_e (-k),c_o^\dag(k), c_o(-k) \right)~,
\label{spinor2}
\end{equation}
and
\begin{equation}
\label{Hk-ti}
  \mathcal{H}(k) = \left(%
\begin{array}{cc}
  \frac12 \hat{M}_e & \hat{Q} \\
  \hat{Q}^\dag  & \frac12 \hat{M}_o \\
\end{array}%
\right)~.
\end{equation}
Here
\begin{widetext}
\begin{equation}
\label{Meo}
  \hat{M}_{e/o} \equiv  \left(%
\begin{array}{cc}
  (1-\gamma \pm \delta) +(1+ \gamma \mp \delta) \cos 2k & -i(1+\gamma \mp \delta )\sin 2k \\
   i(1+\gamma \mp \delta )\sin 2k & -(1-\gamma \pm \delta) -(1+ \gamma \mp \delta) \cos 2k \\
\end{array}%
\right)~,
\end{equation}
and
\begin{equation}
\label{Q}
 \hat{Q} \equiv  \left(%
\begin{array}{cc}
 h \cos k + i h_a \sin k & -i h \sin k -h_a \cos k \\
  -i h \sin k +h_a \cos k & -h \cos k - i h_a \sin k \\
\end{array}%
\right)~,
\end{equation}
\end{widetext}
In the limit $h=h_a=0$ the off-diagonal block $\hat{Q} \to 0$. Then the averaging in the even and odd sectors decouples, and the correlation
functions of the even/odd Majorana operators can be evaluated independently. One can check from the above formulas that these quantities can be calculated from the Toeplitz matrices with the generating functions of their elements given by the standard expressions known from the solution of the Ising chain in transverse field.\cite{McCoyBook} Unfortunately, when $\hat{Q} \neq 0$ such technical simplification is no longer available.

A unitary transformation
\begin{equation}
\label{U}
\hat U=
\left(
  \begin{array}{cccc}
    1 & 0 & 0 & 0 \\
    0 & 0 & 1 & 0 \\
    0 & 1 & 0 & 0\\
    0 & 0 & 0 & 1 \\
  \end{array}
\right)
\end{equation}
brings the spinor \eqref{spinor2} into the new form
\begin{equation}
 \tilde  \psi_{k}^{\dag} = \hat U\psi_{k}^{\dag}=\left(c_e^{\dag}(k),
  c_o^{\dag}(k),c_e(-k), c_o(-k)\right)~.
\label{spinor3}
\end{equation}
The transformed Hamiltonian matrix \eqref{Hk-ti}
\begin{equation}
  \label{Hk-ti2}
   \tilde{\mathcal{H}}(k) = \hat U   \mathcal{H}(k) \hat U
\end{equation}
is brought to the form \eqref{Hk} with the new matrices $\hat A$ and $\hat B$.
The rest can be done along the lines of the analysis presented in the main text. We will
not however elaborate further and present more results on this formalism, since it did
not give us a clear advantage in dealing with the case $h,h_a \neq 0$.
%
%
%%%%%%%%%%%%%%%%%%%%%%%%%%%%%%%%%%%%%%%%%%%%%%%%%%%%%%%%%%%%%%%%%%%%%%%%%%%%%%
\section{String operators $O_x$, $O_y$ and their correlation functions }\label{AppB}
%%%%%%%%%%%%%%%%%%%%%%%%%%%%%%%%%%%%%%%%%%%%%%%%%%%%%%%%%%%%%%%%%%%%%%%%%%%%%%
%
%
In addition to the string \eqref{Ox} one can define the
even and odd string operators:\cite{GYC2018}
\begin{eqnarray}
  \label{Oxe}
  O_{x,e}(m) &\equiv&   \prod_{l=1}^{m} \big[ i b_{2l-1} a_{2l} \big]=
 \prod_{l=1}^{m} \big[ i \tilde b_{2l-2} \tilde a_{2l} \big] \nonumber \\
 &=&  \prod_{n=1}^{2m} \sigma_n^x = \tau_0^x \tau_{2m}^x  \\
  \label{Oxo}
  O_{x,o}(m) &\equiv&  \prod_{l=1}^{m} \big[ i b_{2l} a_{2l+1} \big]=
   \prod_{l=1}^{m} \big[ i \tilde b_{2l-1} \tilde a_{2l+1} \big] \nonumber \\
   &=&   \prod_{n=2}^{2m+1} \sigma_n^x = \tau_1^x \tau_{2m+1}^x
\end{eqnarray}
In the above formulas we used the second auxiliary set of the Majorana operators (distinguished by tildes) defined by
\eqref{Maj-ti}. These operators are very insightful for dealing with the even and odd sectors of the
Majorana Hamiltonian \eqref{Hsum-ti}. The string operators \eqref{Oxe} and \eqref{Oxo} are also presented in terms of
the dual spins $\tau$ using \eqref{XX} and \eqref{XX-ti}.
The Majorana string operator \eqref{Ox} is related to the  above operators as
\begin{equation}
\label{OxRel}
  O_{x,e}(m)O_{x,o}(m)=O_x(2m+1)
\end{equation}
The even/odd SOPs $\mathcal{O}_{x,\sharp}$ ($\sharp =e,o$) are introduced as
\begin{equation}
\label{SOPeo}
 \mathcal{O}^2_{x,\sharp} =\lim_{(n-m) \to \infty} |\langle O_{x,\sharp}(n) O_{x,\sharp}(m)  \rangle |~.
\end{equation}
From \eqref{XX}, \eqref{Oxe}, \eqref{Oxo} one can establish an important
relation \cite{GT2017,GYC2018} between the nonlocal even/odd SOPs and the local dual sublattice magnetizations of the
dual $\tau$ spins:
\begin{equation}
\mathcal{O}_{x,e/o} ^2 = \lim_{(R-L) \rightarrow\infty} \left<\tau_{L}^{x}\tau_{R}^{x}\right>~,
\label{SOPxeo}
\end{equation}
if the parity of $L$ and $R$ is chosen in agreement with \eqref{LMpm}. Another operator's identity
\begin{eqnarray}
  \sigma_{2m+1}^x \sigma_{2n+1}^x
   &=&  O_{x,e}(m) O_{x,e}(n) O_{x,o}(m) O_{x,o}(n) \nonumber \\
   &=&  \tau_{2m}^x \tau_{2n}^x  \tau_{2m+1}^x \tau_{2n+1}^x
  \label{SigOtau}
\end{eqnarray}
allows to establish an important physical property: the spontaneous magnetization of ``original" spins is
due to overlap of the even and odd SOPs.\cite{NoteSig-O-tau}  In the absence of coupling between the even and odd
sectors of the Hamiltonian \eqref{Hmix} when $h=h_a=0$, the above identity results in \cite{GT2017}
\begin{equation}
\label{mxO}
  m_x= \mathcal{O}_{x,e} \mathcal{O}_{x,o}
\end{equation}
and Eq.~\eqref{dg} as a consequence. When $H_{mix} \neq 0$ the factorization of the contributions from the even and
odd sectors does not occur.

The even and odd SOPs are numerically calculated from the determinant of the ordinary $N \times N$ Toeplitz matrix:
\begin{widetext}
\begin{equation}
\label{OxeoDet}
  \left|
    \begin{array}{cccc}
     G^\mp(-1) & G^\mp(1) & \ldots & G^\mp(1-2N) \\[0.2cm]
      G^\mp(1) & G^\mp(-1) & \ldots &  G^\mp (3-2N) \\[0.25cm]
      \vdots & \ddots & \ddots & \vdots \\[0.25cm]
      G^\mp (2N-3) & G^\mp(2N-5) & \ldots &  G^\mp (-1) \\
    \end{array}
  \right| \xrightarrow[N \to \infty]{~} \mathcal{O}_{x,e/o}^2
\end{equation}
\end{widetext}

To probe additional nonlocal orders we utilize another pair of string operators:\cite{GYC2018}
\begin{eqnarray}
  \label{Oye}
  O_{y,e}(m) &\equiv& \prod_{l=1}^{m} \big[ -i a_{2l} b_{2l+1} \big] =
   \prod_{l=1}^{m} \big[ i \tilde b_{2l} \tilde a_{2l} \big] \nonumber \\
 &=& \prod_{n=2}^{2m+1} \sigma_n^y = \prod_{l=1}^{m} \tau_{2l}^z~,   \\
  \label{Oyo}
  O_{y,o}(m) &\equiv& \prod_{l=1}^{m} \big[ -i a_{2l-1} b_{2l} \big] =
  \prod_{l=1}^{m} \big[ i \tilde b_{2l-1} \tilde a_{2l-1} \big] \nonumber \\
 &=& \prod_{n=1}^{2m} \sigma_n^y =   \prod_{l=1}^{m} \tau_{2l-1}^z ~.
\end{eqnarray}
The corresponding SOPs are defined similarly to \eqref{SOPeo} \cite{NoteOy}
and are numerically calculated from the following $N \times N$ Toeplitz determinant:
\begin{widetext}
\begin{equation}
\label{OyeoDet}
  \left|
    \begin{array}{cccc}
     G^\mp(1) & G^\mp(-1) & \ldots & G^\mp(3-2N) \\[0.2cm]
      G^\mp(3) & G^\mp(1) & \ldots &  G^\mp (5-2N) \\[0.25cm]
      \vdots & \ddots & \ddots & \vdots \\[0.25cm]
      G^\mp (2N-1) & G^\mp(2N-3) & \ldots &  G^\mp (1) \\
    \end{array}
  \right| \xrightarrow[N \to \infty]{~} \mathcal{O}_{y,e/o}^2
\end{equation}
One can establish relation \cite{GT2017,GYC2018,NoteOy} between the nonlocal SOPs and the sublattice magnetization of the
dual spins:
\begin{equation}
\mathcal{O}_{y,e/o} ^2 = \lim_{(R-L) \rightarrow\infty} \left<\tau_{L}^{y}\tau_{R}^{y}\right>~.
\label{SOPyeo}
\end{equation}
Similarly to the results of subsection B, the spontaneous magnetization $m_y$ of the original spins $\sigma$ can be determined
from the correlation function of the string  operator $O_y(2m+1)= O_{y,e}(m)O_{y,o}(m)$, cf. \eqref{OxMx}.

In the limit $h=h_a=0$ the above determinants can be evaluated exactly by the standard
technique, \cite{McCoyBook} reproducing the earlier results for nonvanishing $\mathcal{O}_{x/y,e/o}$ obtained from duality mappings.\cite{GT2017}
When $h \neq 0$ and/or  $h_a \neq 0$ we were unable to derive analytical results for asymptotics of these Toeplitz determinants.
Numerical results show that in the presence of fields all SOPs $\mathcal{O}_{x/y,e/o}$ die off in the thermodynamic limit $N \to \infty$.
So the only nonvanishing SOP is $\mathcal{O}_{z}$ discussed in the main text. Similarly to the results \eqref{SOPxeo} and \eqref{SOPyeo} yielding
a simple local dual interpretation of the SOPs  $\mathcal{O}_x$  and $\mathcal{O}_y$, the duality transformation (\ref{sigmaX},\ref{sigmaY}) with the interchange $x \leftrightarrow z$ brings the SOP $\mathcal{O}_z$ to the long-ranged order of the dual $\tau^z$ spins. We emphasize this possibility
to map the string order onto a local order in terms of some judiciously chosen dual variables. We will not go into mathematical details for
the case of $\mathcal{O}_z$, since it is not useful at this point for getting analytical results.
\end{widetext}

\end{appendix}
%%%%%%%%%%%%%%%%%%%%%%%%%%%%%%%%%%%%%%%%%%%%%%%%%%%%%%%%%%%%%%%%%%%%%%%%%%%%%%
%%%%%%%%%%%%%%%%%%%%%%%%%%%%%%%%%%%%%%%%%%%%%%%%%%%%%%%%%%%%%%%%%%%%%%%%%%%%%%

%xxxxxxxxxxxxxxxxxxxxxxxxxxxxxxxxxxxxxxxxxxxxxxxxxxxxxxxxxxxxxxxxxxxxxxxxxxxxxx
% REFERENCES
%xxxxxxxxxxxxxxxxxxxxxxxxxxxxxxxxxxxxxxxxxxxxxxxxxxxxxxxxxxxxxxxxxxxxxxxxxxxxxx
%

%
%xxxxxxxxxxxxxxxxxxxxxxxxxxxxxxxxxxxxxxxxxxxxxxxxxxxxxxxxxxxxxxxxxxxxxxxxxxxxxx
%

\begin{thebibliography}{}
\bibitem{Note}
The exception is the IC critical line when $\gamma = 0$.
It belongs to the different universality class
of the conformal charge $c=1$. (The $2D$ Ising belongs to the $c=\frac12$ class).
\bibitem{NoteSig-O-tau}
A similar identity can be derived for the spins $\sigma$ on even sites:
\begin{eqnarray}
  \sigma_{2m}^x \sigma_{2n}^x
   &=&  O_{x,e}(m) O_{x,e}(n) O_{x,o}(m-1) O_{x,o}(n-1) \nonumber \\
  &=&  \tau_{2m}^x \tau_{2n}^x  \tau_{2m-1}^x \tau_{2n-1}^x \nonumber
  \label{SigOtauEven}
\end{eqnarray}
\end{thebibliography}
\end{document}